\documentclass[baaa]{baaa}

\usepackage[pdftex]{hyperref}
\usepackage{subfigure}
\usepackage{natbib}
\usepackage{helvet,soul}
\usepackage[font=small]{caption}

\contriblanguage{1}

\contribtype{6}

\thematicarea{3}

\received{\ldots}
\accepted{\ldots}

\title{Stars, bow shocks, and gamma-ray sources}

\titlerunning{Stars, bow shocks and gamma-ray source}

\author{
P. Benaglia\inst{1}
}

\authorrunning{Benaglia}

\contact{pben.radio@gmail.com}

\institute{  
Instituto Argentino de Radioastronom\'ia, CONICET--CICPBA--UNLP, Argentina
}

\resumen{Aquí presento los resultados de mis investigaciones realizadas en el último par de décadas, focalizados mayoritariamente en los campos de la astrofísica estelar, el medio interestelar y las fuentes de altas energías. Los mismos han sido obtenidos principalmente a partir de observaciones interferométricas dedicadas en la banda de radio a bajas frecuencias del espectro electromagnético.}

\abstract{Here I present a somewhat personal review of the results of my research carried out in the last couple of decades, mostly focused in the fields of stellar astrophysics, the interstellar medium, and the high energy sources. They have been obtained mainly from dedicated interferometric observations in the radio band at low frequencies of the electromagnetic spectrum.}


\begin{document}

\maketitle
\section{Introduction: some biographical notes}\label{intro}

In 1997, during a postdoctoral stay abroad, I came across with an article that started the main research project { transversal} to my career -- and in development-- until now: understanding the emission of radio stars. \citet{leitherer1995} published the results of a pilot project: the first detections of a few massive early-type stars (METS) in the southern sky,  
with the amazing Australia Telescope Compact Array (ATCA). ATCA consists of six 22~m diameter dishes, originally lined up east-west, but  with only 15 baselines covering a huge field area unseen by instruments in the northern hemisphere, down to an arcsecond angular resolution. All sources with declinations below $-40$~deg were waiting to be surveyed with the highest angular resolution available at that time. ATCA calibration and analysis software was also a good news at least in terms of practicality and friendliness, compared to the powerful Very Large Array, whose results, fair to quote, since 1981 have revolutionized the cm-wavelength data range. 

With the basics of radio interferometry learned after a hard year as an assistant researcher at the Very Large Array (VLA, Socorro, NM), the studies started in teaming first with Baerbel Koribalski for ATCA observations of METS \citep{benagliamdot200}. Over the years,  VLA and Giant Metrewave Radio Telescope (GMRT) observations have also been made, extensively, in this line of research. And other phenomena, mostly also related to stars, completed the set of studies performed.
In this article, I present the motivations that triggered the studies, and some of the discoveries.
 

\section{Stellar objects in  general}\label{sec:SOs}

\begin{figure*}[!t]
\centering
\includegraphics[width=16.5cm]{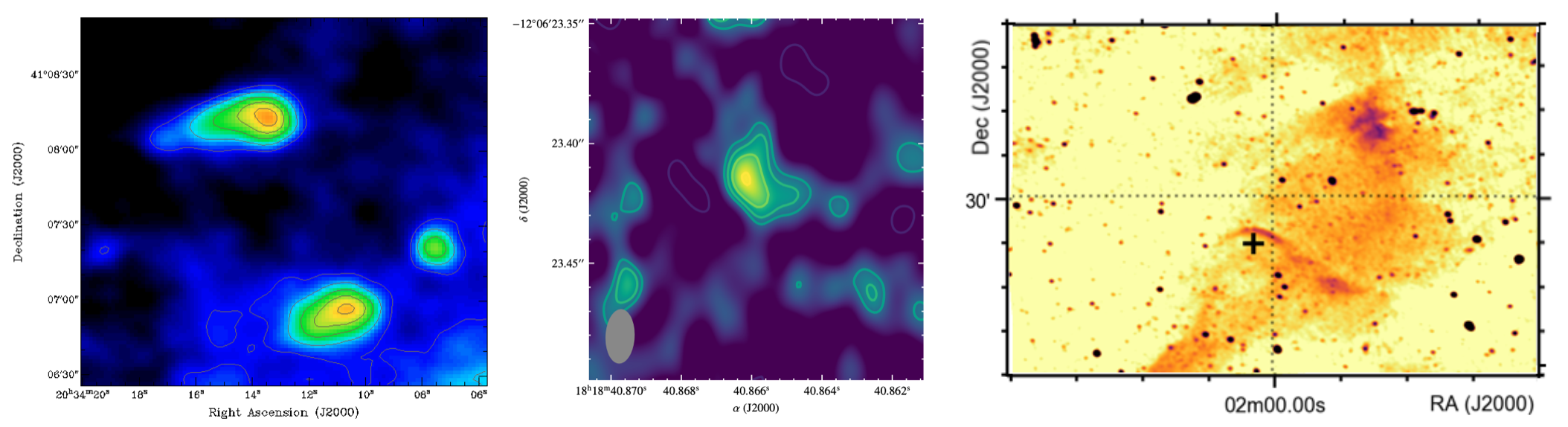}
\caption{Examples of stellar objects. \emph{Left panel}: Proplyd-like source; continuum emission at 610~MHz, taken with the GMRT (PI: Benaglia), of the Tadpole (northern source), see text and also \citet{isequilla2019}. \emph{Central panel}: colliding-wind region of the binary system HD\,168112; continuum emission at 1.67~GHz, taken with the European VLBI Network \citep{debecker2024}. \emph{Right panel}: Vela~X-1 bow shock, as seen with MeerKAT in L-band \citep[adapted from Fig.~1 of][]{vandenmeer2022}.}
\label{fig:stellarobjects}
\end{figure*}

The target sources of the studies can be grouped under the collective term ``stellar objects'': stars at different stages of evolution, from protostellar cores to those in the pre-supernova phase, e.g. Fig.~\ref{fig:stellarobjects}.

A good example of the former is IRAS~16353--4643. We carried out observations with ATCA in the radio continuum from 1.4 to 20~GHz, and with ESO-NTT, including photometry and spectroscopy, in the IR bands JHK$_{\rm s}$. The nature of the source could be then identified as a new protostellar cluster with low and high-mass young stellar objects (YSOs) and non-thermal (NT) emission, probably arising in an outflow \citep{benagliairas2010}.

Protostellar cores or clusters, when close enough to sources of intense UV flux, become externally ionized. Such structures, with a cometary tail opposite the ionizing source, are called protoplanetary disk-like sources, or simply proplyds. In the framework of a Ph.D. thesis, a number of proplyds towards the Cyg~OB2 association were characterized using data below 1~GHz \citep{isequilla2019}, including the famous Tadpole \citep{sahai2012}, and the ionizing sources in some cases could be  proposed. For some, negative spectral indices ($\alpha$, where $S \propto \nu^\alpha$) were measured. However, the scenario of synchrotron emission could not be conclusively concluded. 

Protostellar jets are a common phenomenon during the star-formation phase; their interaction with the interstellar medium can lead to strong shocks, particle acceleration, and subsequent synchrotron radiation. By way of confirmation of this latter, \citet{carrasco2010} found highly polarized emission from the Herbig Haro objects HH80--81.
In \citet{lopezsantiago2013} hard X-ray emission was reported, as another indication of non-thermal processes. 

Since those early ATCA observations, most of the work has been devoted to stellar winds of massive stars and their interaction with the environment, strongly supported by radio interferometric observations, with very large arrays but also very large baseline interferometers. Those two sub-fields of research are discussed in separate sections (\ref{sect:cwb} and \ref{sect:bowshocks}).

\section{The problem with the high energy sources}\label{sect:HEsources}

Are stellar objects gamma-ray emitters? The personal quest to answer this question began with the advent of all-sky gamma-ray observations obtained with the Energetic Gamma-Ray Experiment Telescope (EGRET) aboard the Compton Gamma-Ray Observatory. The 3rd EGRET catalog listed 271 sources, detected in the spectral range 30~MeV -- 20~GeV.
170 out of the 271 sources could not be at first clearly identified with known objects \citep{hartman1999}. The nominal PSF of the images was $\sim 0.5$~deg.

On June 11, 2008, the Fermi satellite was launched, carrying as its main instrument the Large Area Telescope, which operates between 1~keV and 300~GeV. Thereafter, the Fermi collaboration has released a catalog about every two years. The last fully documented release, 4FGL--DR3 \citep[4th Fermi catalog 3rd data release,][]{abdollahi2022}, contains nearly 6700 point sources after processing data from 12~yr of observations. The note of \citet{ballet2023} lists first results on 14-yr data (DR4, 7194 entries, incremental catalog). 

At TeV energies and beyond, the Large High Altitude Air Shower Observatory (LHAASO) collaboration recently published a catalog of 90 sources covering the declination interval [$-$20, $+80$] \citep{lhaasocat2023}. However, the large size of the sources (up to 2~deg) precludes  conclusive identification and correlation with individual objects in many cases. 

\begin{figure*}[!t]
\centering
\includegraphics[width=16.5cm]{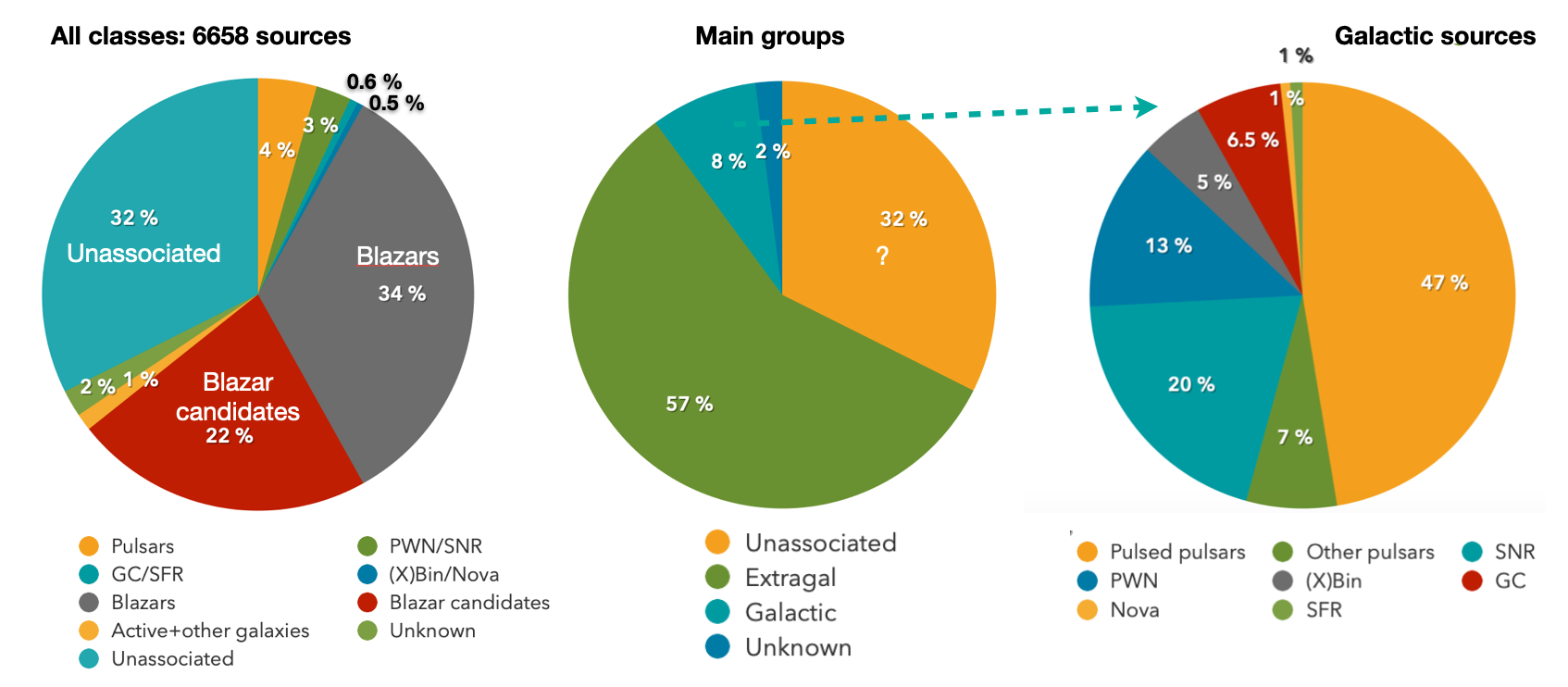}
\caption{{Distribution} of the sources of the 4FGL--DR3 catalog.  GC = globular cluster; SFR: =  star forming regions; PWN = pulsar wind nebula; SNR = supernova remnants; (X)Bin = binaries or X-ray binaries. Extragal = extragalactic sources. \emph{Left panel:} Percentage of nine groups/classes. \emph{Central panel:} Percentage of four main groups relevant to our work. \emph{Right panel:} Percentages covered by galactic source classes.}
\label{fig:piecharts}
\end{figure*}

Each time a catalog goes public, much effort is expended to confirm or reveal the nature of the new sources. In general, a large part of them have no clear counterpart at lower energies. 
Figure~\ref{fig:piecharts} shows the situation in the case of the 4FGL-DR3. Sources are marked as `identified' if spectral, timing, or other studies have been performed and confirm their nature; as `associated' if the positional coincidence between a plausible gamma-ray producing object and a Fermi source is statistically unlikely to have occurred by chance; and `unknown' are those associated only by the likelihood ratio method from large radio and X-ray surveys. 

It can be appreciated from Fig.~\ref{fig:piecharts} that one third of 4FGL-DR3 sources were unassociated. If one divides the sources in four groups, --unassociated, extragalactic, galactic and unknown --, the second class is the more numerous: their classes are easier to be identified/associated. The same happens with galactic pulsars. 

On the contrary, the number of Fermi-LAT sources related to massive stellar systems was very low. They were 9 of them, listed in Table~\ref{tabla1}. 

In short, of the 4FGL catalog, roughly 10\,\% are galactic in nature, 2\,\% are of unknown nature, and 30\,\% (many hundred of sources) have no association/identification.
The goals of that initial search performed using the 3rd EGRET catalog remain the same today: to search for (mainly stellar) objects of known types as counterparts of gamma-ray sources and provide support for models; study new populations that may be counterparts, and apply the methods and results to explain upcoming unassociated/unidentified sources to be detected by the Cherenkov Telescope Array\footnote{Common interval between CTA and Fermi-LAT: 20--300~GeV; CTA PSF = $4.2'$; estimated population with CTA: hundreds of galactic sources will be discovered with the ``CTA Galactic Plane Survey''.}.

\begin{table}[!h]
\centering
\caption{Stellar sources of the 4FGL-DR4 catalog.}
\begin{tabular}{l c c}
\hline\hline\noalign{\smallskip}
Source & type  &  Reference\\
\hline\noalign{\smallskip}
Eta Carinae    & Identified & Ps2016 \\
Gamma2 Velorum & Associated & MD2021\\
V918 Sco       &  Associated & A2022 \\
4FGL J1408.6-2927& Associated & A2022 \\
XMMU J083850.3-282756 & Associated & A2022\\
CRTS J052316.9-252737&  Associated & A2022\\
Kleinmann star& Associated & A2022\\
4FGL J0935.3+0901&  Associated & A2022\\
1SXPS J021210.6+532136& Associated & A2022\\
\hline
\end{tabular}
P2016: \citet{pshirkov2016}, MD2021: \citet{martidevesa2021}, A2022: \citet{abdollahi2022}
\label{tabla1}
\end{table}

\section{Colliding-wind stellar systems}
\label{sect:cwb}

Massive, early-type stars are characterized by effective temperatures of tens of K, luminosities of orders of magnitude that of the Sun, and strong winds that strip their outer layers while they are still on the main sequence, depositing matter and momentum in the environment. OB stars above 8 solar masses and their evolved Wolf-Rayet stars are METS.
Radio observations provide a way to derive the mass loss rate, a crucial parameter in stellar evolution studies and the initial mass function. In principle, the winds are sources of thermal (free-free) radiation. 

But METS are usually found in binary or multiple systems, with similar types, coeval. The interaction of their stellar winds is capable of accelerating particles to relativistic energies. Under the wind plasma conditions (strong UV field from the star, magnetic fields), non-thermal -synchrotron- emission is expected. \citet{eichler1993} published a first model to explain how colliding wind binaries (CWB) could be sources of gamma-ray emission.

Since the early days of the VLA (opening: 1981), dozens of METS have been observed and many have been detected, with spectral indices consistent with thermal, non-thermal or combined radio emission \citep[][etc]{bieging1989,scuderi1998,benaglia2006,benaglia2007}. 

\begin{figure}[!t]
\centering
\includegraphics[width=7.5cm]{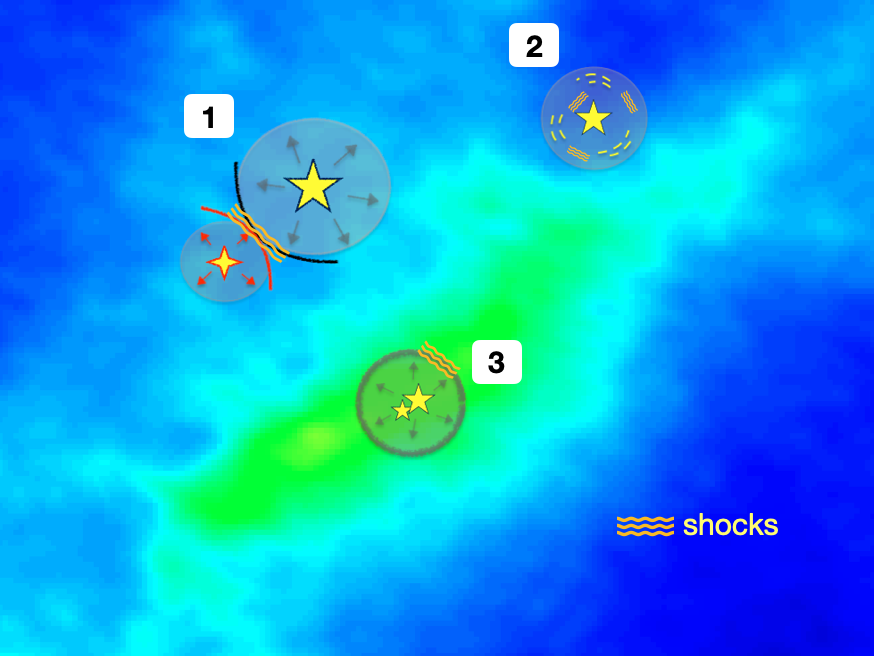}
\caption{Particle acceleration scenarios considered back in 2000 related to massive stars. 1: colliding winds; 2: single stellar wind; 3: interaction of ISM and terminal wind.}
\label{fig:scenarios}
\end{figure}

With the release of the EGRET catalogs, we started looking for counterparts in the massive star collection. In \citet{romero1999} we showed that an  association by chance was high for WR and Of stars (those with more intense winds). In \citet{benagliarom2003} we described the main scenarios for the acceleration of particles in METS: shocks in the colliding wind regions of binaries, in single star winds, and in the stellar wind terminal region where it hits a rather dense ISM/cloud (see Fig.~\ref{fig:scenarios}). Three processes for gamma-ray production were considered: inverse Compton scattering of relativistic electrons in an intense stellar UV photon field, relativistic bremsstrahlung of electrons idem in relevant matter fields, and interaction between field and relativistic protons producing neutral pion decay to gamma rays. The calculations were applied to emblematic WR+O systems, WR~140, WR~146, and WR~147. Although we worked with many assumptions and a first order approximation formulae approach, the value of the paper was to obtain orders of magnitude of the main contributions in each case.  
\citet{benagliacyg2001} and \citet{benaglia2005} discussed, for two EGRET sources, the probability that the systems Cyg~OB2~\#5 and WR~21a can contribute; both IAR single dish and ATCA data were collected to do so.

\begin{figure*}[!t]
\centering
\includegraphics[width=16cm]{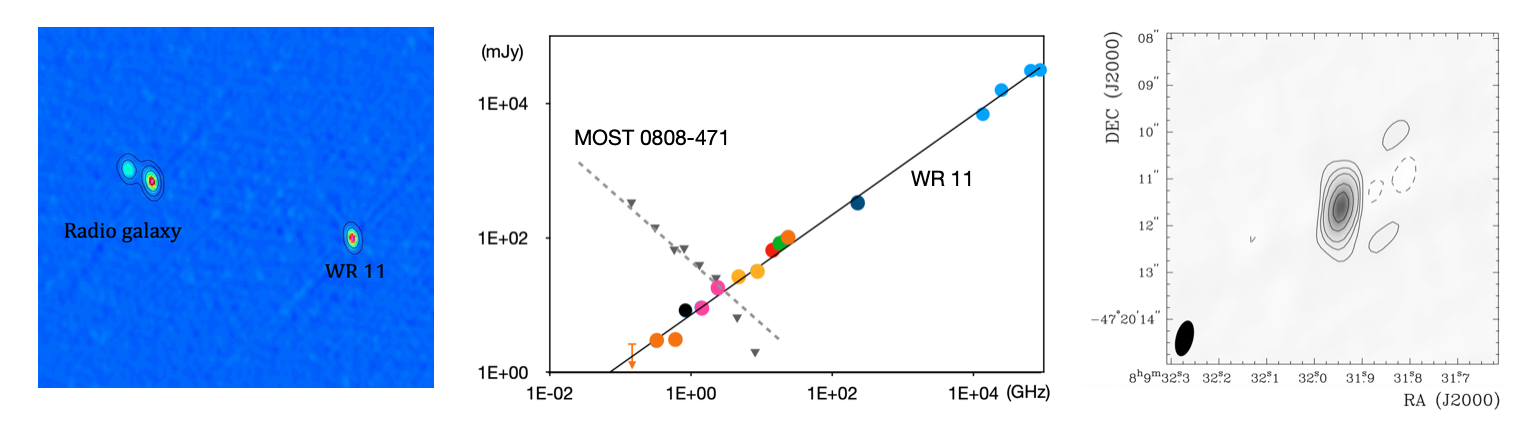}
\caption{Results from studies of Gamma$^2$Vel (WR~11). \emph{Left panel}: Sources inside the Fermi excess measured by \citet{pshirkov2016}, in continuum emission at 4.8~GHz (ATCA data). Taken from the ATNF Picture of the day, \url{https://www.atnf.csiro.au/ATNF-DailyImage/archive/2016/04-Apr-2016.html}.  \emph{Central panel}: Spectra of the sources at the left panel; MOST~0808-471 is identified as a radio galaxy; flux density values given in \citet{benaglia2019}. \emph{Right panel}: Continuum emission at 22~GHz from ATCA data (C1616, PI: Benaglia).}
\label{fig:wr11}
\end{figure*}

The 4FGL-DR4 presents the system Eta~Car (HD~93308) as the only CWB identified with a Fermi source, also detected by the instruments Astrorivelatore Gamma ad Imagini Leggero (AGILE) and High Energy Stereoscopic System (H.E.S.S.). Nevertheless, it is a thermal radio emitter, widely observed from centimeter to millimeter wavelengths. Eta~Car is composed of an LVB and an OB star. The wind of the LVB is very dense ($dM/dt > 10^{-4}$~M$_{\sun}$~yr$^{-1}$) and the OB has a very high terminal velocity ($\sim 3000$~km~s$^{-1}$). 

The Gamma$^2$Velorum system (WR~11) is the only one among METS associated with a Fermi source. It is the nearest colliding binary system ($\sim 350$~pc). We have studied it carrying out several radio experiments. With ATCA at 22~GHz (proposal C1616, PI: Benaglia) we detected no clumps in the wind, neither with the miriad \citep{sault1995} package, nor with the difmap \citep{shepherd1997} one  (S. Dougherty, private communication). No variations in the light curve along the 2-yr interval, from archival ATCA data, were found \citep{benaglia2016}. When \citet{pshirkov2016} reported a 6-$\sigma$ Fermi excess over this system, we implemented a multiband campaign to complement its spectrum from 150~MHz up to the IR range \citep{benaglia2019}. We found a thermal spectral index throughout, and a non-thermal source (MOST~0808--471, a radio galaxy) nearby, within the Fermi ellipse (see Fig.~\ref{fig:wr11}).

A third CWB that we have studied extensively is HD\,93129A \citep[see][and references therein]{benaglia2015}. It is an (O2If*+O2If*) system with a non-thermal spectral index. We performed VLBI observations to detect its colliding wind region, with the Australia Long Baseline Array; we needed to use a first run to look for appropriate calibrators.  It shows no associated gamma-ray emission.\\

Why are there so few detections of massive binaries in the gamma-ray range?

\begin{itemize}

\item Some of these systems are quite eccentric, and the conditions for efficient particle acceleration and gamma-ray production may manifest only occasionally; conditions vary along the orbit, with optimal windows for observations occurring at different times for different wavelengths. 

\item Convection effects can carry away the protons, preventing hadronic interactions from contributing significantly.

\item Current predictions \citep{delpalacio2016,reitberger2017,pittard2021} tend to show that IC emission is lower than expected and that synchrotron losses may be higher than previously thought (e.g. high B-field, tight systems).

\item In addition, the non-thermal radiation can be suppressed by the high densities of thermal gas and the absorbing photon fields.

\item TeV emission requires extremely efficient acceleration.

\item Crucial for the modeling, stellar and orbital parameters are usually not well known; mass loss rate, wind terminal velocity, wind region size, inclination, spectral types, clumping, even distance, stellar temperature, chemical composition, surface magnetic field, etc.

\end{itemize}

\section{Stellar bow shocks}\label{sect:bowshocks}

\begin{figure*}[!t]
\centering
\includegraphics[width=16cm]{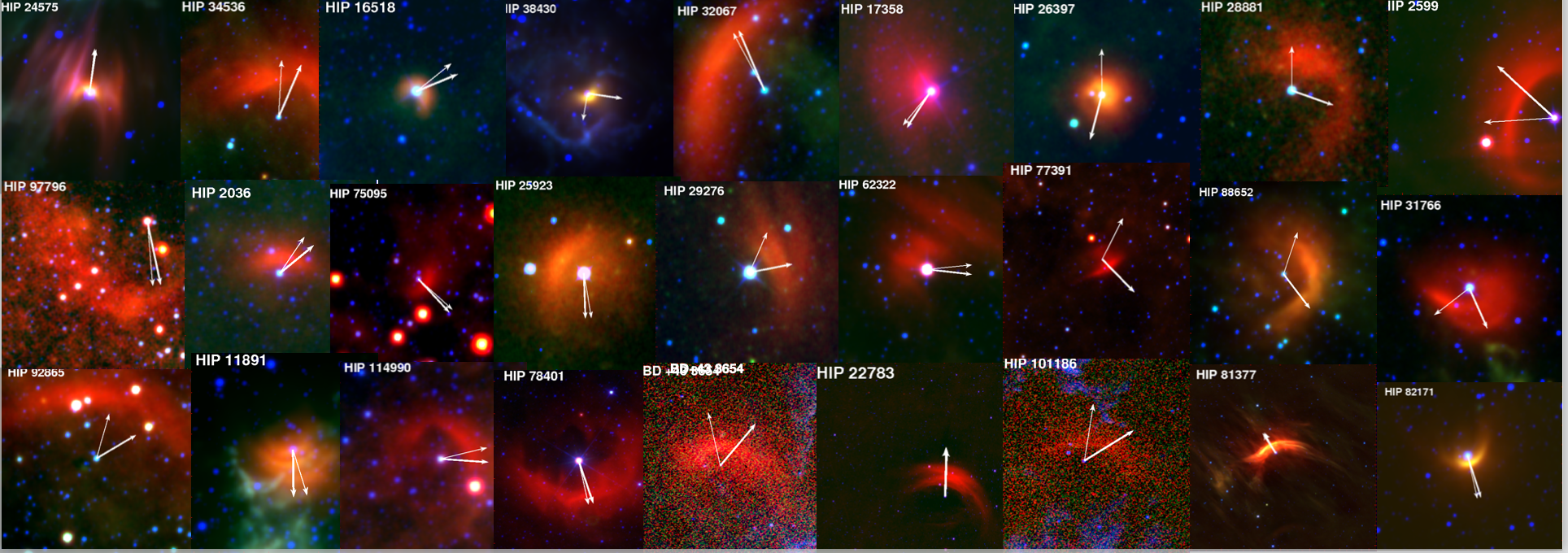}
\caption{Patchwork made with bow shocks from E-BOSS \citep[see][for full details]{peri2015}.}
\label{fig:eboss}
\end{figure*}

The stellar bow shocks (SBS) are formed when a runaway star, with a powerful wind, moves at high velocity above the surrounding medium, sweeping and heating that medium. If supersonic speeds, shocks will be produced.

SBSs are arcuate structures of stacked material, in the same direction as the stellar velocity vector. The stellar winds are confined by the ISM ram pressure. The distance to the star is given by momentum balance between the two media. The shocked stellar radiation heats the accumulated dust, which re-radiates as MIR-FIR excess. One of the first systematic search for SBS was performed by \citet{noriega1997}, who identified 21 candidates, through HIRES IRAS data images. 

Another {landmark} result in my career appeared in 2007: \citet{comeron2007} reported and described a bow shock feature using Midcourse Space eXperiment (MSX) IR data, and pointing to the star BD$+43^0$~3654 as the driver; this bow shock would later be named EB27. The star has spectral classification of O4~If, very high mass loss, terminal and peculiar velocities, and was proposed to be coming from the direction of Cyg~OB2. 
This massive, luminous runaway was an ideal source to find if particles could be accelerated at the shocks.

A fairly straightforward way to test the hypothesis was to observe the IR feature in two low frequency radio bands, where non-thermal emission would be prominent. A spectral index map of EB27 was obtained using VLA data, with values down to $\alpha = -0.45$ \citep{benaglialet2010}. 
{We reasoned that if the following three conditions hold:} (i) the MSX feature was the stellar bow shock, (ii) the radio emission detected with the VLA was primarily produced in the bow shock, and (iii) the negative spectral index was evidence for NT emission, then SBSs could be HE sources. 
 
This result was followed by various models \citep[e.g.][]{delpalacio2018} to explain the {synchrotron} radiation from a SBS, calculate the spectral energy distribution, and predict the high-energy emission. 

Another Ph.D. project consisted of compiling a catalog of SBSs \citep{peri2015} close to runaway stars, the Extensive stellar BOw Shock Survey or E-BOSS, and characterizing its members (see Fig.~\ref{fig:eboss}). 
Low-frequency radio emission from archives were found for five of the E-BOSS SBS. However, systematic searches of correlations with high-energy (HE) counterparts (Fermi, HESS) were unsuccessful \citep{schulz2014,hesscoll2018}.

A second round on EB27 involved deeper, and full polarimetric observations with the Jansky VLA, to search for conclusive evidence of synchrotron emission. The negative spectral index at the SBS core was confirmed. 
Puzzling, the fractional polarization remained $<$ 0.5\%. This could indicate a turbulent magnetic field, magnetic reconnection as the acceleration mechanism -- instead of diffusive shock acceleration--, Faraday rotation through a diffuse and poorly imaged medium due to insufficient short baselines, etc. \citep{benagliabd2021}.


Recently, the full operation of the Australian Square Kilometer Array Pathfinder (ASKAP) and the MeerKAT-Karoo Array Telescope has given new impetus to BSB studies. The classical bow shock of Vela-X1, a runaway HMXB (NS+B0.5Ia), was detected at 1.3~GHz with MeerKAT, and a thermal scenario provided reasonable values for bow-shock density and temperature and ISM density \citep{vandenmeer2022}. 
  
Images from the Rapid ASKAP Continuum Survey (RACS)-low-DR1 were examined at the position of E-BOSS SBSs: 3 bow shocks were identified as strong cases, 3 as possible bow shocks, another one as to be confirmed, and other three were removed from the bow-shock list
\citep{vandenracs2022}.

\citet{moutzouri2022}, using JVLA and Effelsberg data, to add the short-spacings, imaged the Bubble Nebula (NGC~7635); part of this source is a bow shock, probably formed by the runaway star BD$+$60$^0$2522. The radio emission at the SBS has a spectral index clearly indicative of NT emission: finally, a second bow shock with strong indication of NT emission was found.

And with the release of the Gaia-DR3 database \citep{gaia3dr2023}, the systematic search and characterization of runaway stars capable of  producing the BSB has been resumed \citep{carretero2023}. 
 
\section{Large field studies}\label{sec:largestudies}

\subsection{RCW~49 and the Westerlund~2 cluster}
The large sizes of very high-energy (VHE) sources force one to cover large regions (up to several square degrees) when searching for counterparts  at other wavelengths. This means planning mosaicking observations for radio fields, i.e. multiple pointings, and then combining them in a way to unify the noise.

\begin{figure}[!h]
\centering
\includegraphics[width=8cm]{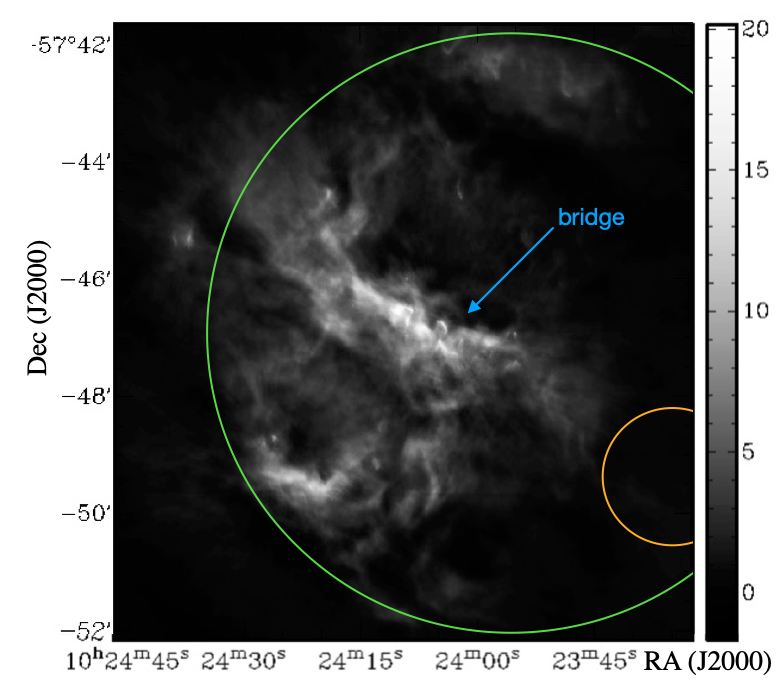}
\caption{The ionized gas of RCW~49, imaged at 5.5~GHz (synthesized beam of $1.9'' \times 1.5''$). The orange contour represents the 2FGL source J1023.5--5749c; the green contour that of HESSS~J1023--575. The `bridge' region is marked. The grayscale units are mJy~beam$^{-1}$.}
\label{fig:rcw49}
\end{figure}

This was the case in a study of the bright H{\sc ii} region RCW~49 and its ionizing cluster Westerlund~2 (Wd2). Figure~\ref{fig:rcw49} shows the probability contours of the sources 2FGL~J1023.5--5749c, and HESS~J1023--575 superimposed on the radio continuum emission. We performed 41 pointings at the most extended array configuration with ATCA, at 5.5~GHz and 9.0~GHz, with 2-GHz bandwidths. The images allowed to detect a bubble with a jet at the {center} of Wd2, and non-thermal emission from a region called the `bridge', which could be related to the production of the gamma-ray sources \citep{benaglia2013}. 

\subsection{The Cygnus constellation core}

\begin{figure*}[!t]
\centering
\includegraphics[width=16cm]{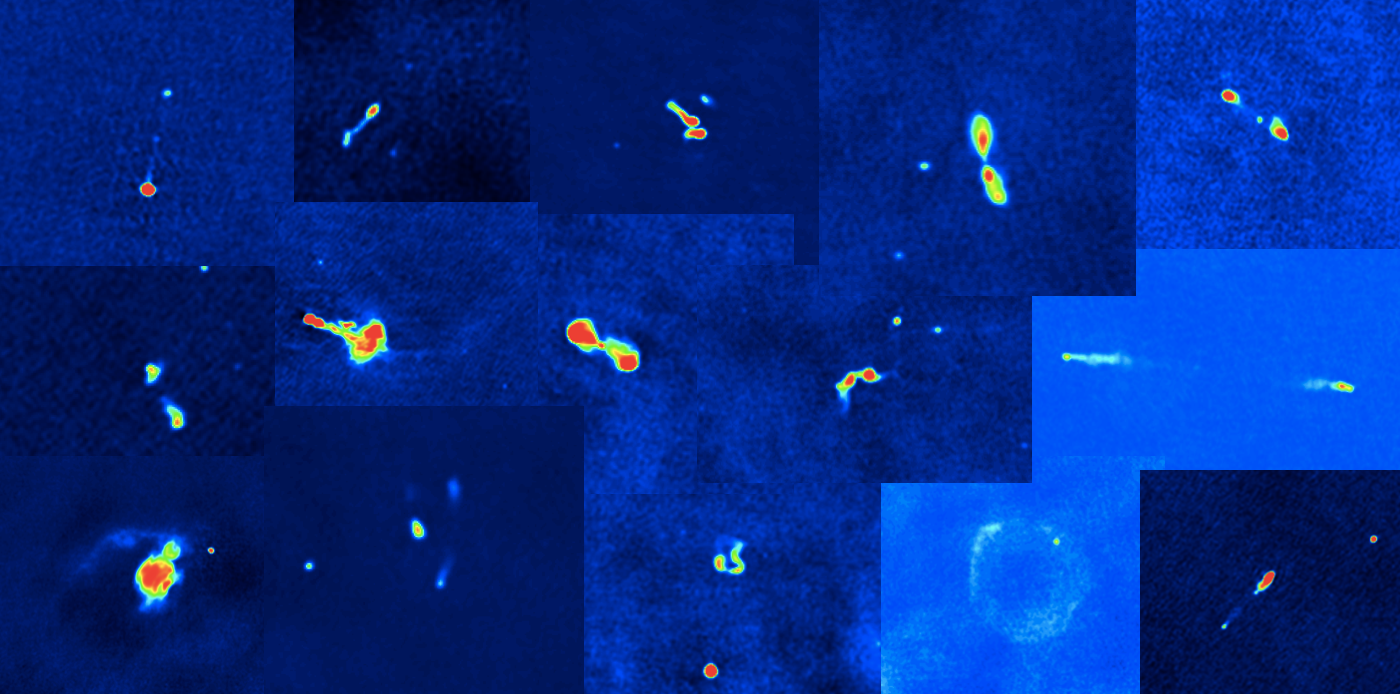}
\caption{Collage of sources in the Cygnus core, emitting in the continuum, at 610~MHz.}
\label{fig:cygnuscollage}
\end{figure*}

The Cygnus Rift, a path of dark patches across many degrees of the northern sky, contains many OB associations and star clusters. Among them is Cyg~OB2, with hundreds of METS and traces of recent star formation. In addition, many TeV and GeV sources, some of them unidentified, have been catalogued \citep[see Fig.~1, and Tables 1 and 2 of][]{benagliacyt2021}.   

{Through extensive work Ishwara Chandra}, we aimed to obtain information or confirmation about the nature of the high-energy sources, by exploiting the radio-gamma connection: a relativistic electron population that can feed radio synchrotron photons and gamma rays. To this end, we completed a project to image Cyg~OB2 and its surroundings at MHz frequencies, covering almost 20~sq.~deg.

The instrument chosen was the GMRT, which operates from 150~MHz to L-band and achieves very high angular resolution. The observing bands were 325 and 610 MHz, with a bandwidth of 32~MHz\footnote{I think one of the best things about being a radio astronomer is to go observing. This prerogative can reach an extreme, like in ATCA, where you even drive the instrument and take care of the observation, with all the responsibility for the result that implies.}. It took us 4 campaigns (2013 to 2017), 185~h of observing, and more than 50 pointings.

Among the most significant results, the first 610~MHz image of such a large area was obtained, together with a 325~MHz image with x5 better sensitivity and angular resolution than previous observations; the rms reached rounded 0.2~mJy~beam$^{-1}$ and 0.5~mJy~beam$^{-1}$, and the angular resolutions were set to $6''$ and $10''$, at 610~MHz and 325~MHz respectively. Thousands of sources were detected {(see Fig.~\ref{fig:cygnuscollage} with some  examples).}


The main catalog was generated by a source detection and characterization pipeline for emission at a level of 7-$\sigma$ or more \citep{benagliacat2020}. It lists 2796 sources at 610~MHz and 933 at 325~MHz, with spectral indices for those at the same position in both bands.

The massive stars/systems of the field were studied; 11 were detected. Many of their full spectra, completed with the values at other frequencies from the literature, showed indications of free-free absorption \citep{benagliamets2020}.



Since the images were constructed with weighting to outline discrete sources, those coinciding with YSOs in the field could be identified \citep[$\sim$30][]{isequilla2020} and the same for the double-lobed sources \citep[$\sim$40][]{saponara2021}. 

The analysis of the radio counterparts superimposed to gamma-ray sources was performed considering radio sources above 3 times local rms. We found 36 radio sources on 10 HE sources, and 11 on VHE sources \citep{benagliamets2020}. A few radio sources were resolved. The first 325-MHz image of the SNR Gamma-Cygni was obtained.
Variability was also checked against previous works.

Eleven radio sources showed non-thermal emission. Most of them, corresponding to eight gamma-ray sources, 
were characterized by a C-factor \citep[flux density/peak brightness,][]{frail2018} $<$1.5, indicating a pulsar candidate. 
Both timing studies to confirm pulses and VLBI data to distinguish resolved high-redshift galaxies from unresolved pulsars are underway.

\subsection{The PACWB project}

Regarding the scenarios shown in Fig.~\ref{fig:scenarios}, not much progress has been accomplished for single METS as non-thermal sources \citep{vanloo2006,anindya2023}. The fact that the binarity/multiplicity of a star can be very tricky to determine aggravates the problem. 

The relationship between a non-thermal spectral index and multiplicity was discussed early in \citet{dougherty2000}. The authors analyzed the radio spectral index information of 25 WR stars and found that 7 of them had non-thermal emission, 5 of which were in massive systems.

\citet{jaenproc2010} compiled the results corresponding to 64 stars with spectral types O to B2 and radio emission, and found that 17 of them presented non-thermal emission, 8 showed hints of NT emission, 18 presented thermal emission, 7 had hints of thermal emission, and for the rest of them (14) the spectral index calculation led to unclear results.

Later, the `Catalog of Particle-accelerating colliding-wind binaries' (PACWBC),  was published \citep{pacwb2013}. It is maintained and updated  at \url{https://www.astro.uliege.be/~debecker/pacwb/}. It hosts about 50 members with radio observational clues of NT emission, plus about 15 members-to-be-confirmed. Clues include a clear non-thermal spectral index, flux density variability along the orbital phase, high  flux density values that cannot be explained by thermal emission alone, especially at low frequencies, point-like emission, gamma-ray associated source, NT X-ray emission, etc. 

For a binary system to show NT radio emission, the winds must be close enough to interact, but not so close for the NT photons produced by the interacting winds can escape the volume filled by the individual thermal winds, where they tend to be absorbed (FFA). This is why we distinguish systems that can accelerate particles (up to relativistic speeds) from those that cannot. 

The PACWBC promoted the initiative PANTERA-stars\footnote{\url{https://www.astro.uliege.be/~debecker/pantera/}} {(leaded by Micha\"el De Becker)}, an international collective to study Particle Acceleration and NT Emission of Radiation in Astrophysics.

The main goals of PANTERA-Stars are to confirm non-thermal emission (spectral index, brightness, variability, polarization), to study the origin of the variability (e.g. related to the orbital phase), to search for new PACWB (deeper observations with new or updated instruments), to analyze the relation between NT emission and binarity/multiplicity, and to search for NT emission in presumed single stars.

\begin{acknowledgement}
P.B. thanks the Jury of the Sérsic Prize of the Asociación Argentina de Astronomía, and the Local Organizing Committee of the 2023 Annual Meeting 
for providing the means to travel to the city of San Juan to participate. P.B. is very grateful to G.E. Romero, B. Koribalski, Ishwara Chandra C.H., J.M. Paredes, J.S. Vink, J. Martí, M. De Becker, A. Tej, A. Noriega-Crespo, and many other colleagues for sharing to a greater or lesser extent their expertise. Some of this material was presented as part of the invited talk at the 9th Fermi Symposium, which took place virtually back in 2021.
\end{acknowledgement}


\bibliographystyle{baaa}
\small
\bibliography{bibliografia-benaglia}
 
\end{document}